\documentclass[aps,prd,superscriptaddress,showpacs]{revtex4}
\usepackage{graphicx}
\usepackage{amsmath}
\usepackage{amsfonts}
\usepackage{amssymb}
\usepackage{latexsym}

\begin{document}
\title{Finite Field-Energy and Interparticle Potential in Logarithmic Electrodynamics}
\author{Patricio Gaete}\email{patricio.gaete@usm.cl}
 \affiliation{Departmento de F\'{i}sica and Centro Cient\'{i}fico-Tecnol\'ogico de Valpara\'{i}so, Universidad T\'{e}cnica Federico Santa Mar\'{i}a, Valpara\'{i}so, Chile}
\author{Jos\'{e} Helay\"{e}l-Neto}\email{helayel@cbpf.br}
\affiliation{Centro Brasileiro de Pesquisas F\'{i}sicas (CBPF), Rio de Janeiro, RJ, Brasil}

\begin{abstract}
We pursue an investigation of Logarithmic Electrodynamics, for which the field-energy of a point-like charge is finite, as it happens in the case of the usual Born-Infeld electrodynamics. We also show that, contrary to the latter, Logarithmic Electrodynamics exhibits the feature of birefringence. Next, we analyze the lowest-order modifications for both Logarithmic Electrodynamics and for its non-commutative version, within the framework of the gauge-invariant path-dependent variables formalism. The calculation shows a long-range correction ($1/r^5$- type) to the Coulomb potential for Logarithmic Electrodynamics. Interestingly enough, for its non-commutative version, the interaction energy is ultraviolet finite. We highlight the role played by the new quantum of length in our analysis.
\end{abstract}
\maketitle

\section{Introduction}

The photon-photon scattering of Quantum Electrodynamics (QED) and its physical consequences such as vacuum birefringence and vacuum dichroism have been of great interest since its earliest days \cite{Adler, Constantini, Biswas, Tommasini, Ferrando, Seco, Kruglov1}. Even though this subject has had a revival after recent results of the PVLAS collaboration \cite{Zavattini, Bregant}, the issue remains as relevant as ever. We also point out that alternative scenarios such as Born-Infeld theory \cite{Born}, millicharged particles \cite{Gies} or axion-like particles \cite{Masso, Gaete1, Gaete2} in order to account for the results reported by the PVLAS collaboration. 

We further note that recently considerable attention has been paid to the study of nonlinear electrodynamics due to its natural emergence from D-brane physics, where the Born-Infeld theory plays a prominent role. In addition to the string interest, nonlinear electrodynamics have also been investigated in the context of gravitational physics. In fact, Hoffman \cite{Hoffman} was the one who first considered the connection between gravity and nonlinear electrodynamics (Born-Infeld theory). In passing we recall that these nonlinear gauge theories are endowed with interesting features, like finite electron self-energy and a regular point charge electric field at the origin. Very recently, in addition to Born-Infeld theory, other types of nonlinear electrodynamics have been studied in the context of black hole physics \cite{Hendi, Zhao, Olea, Habib}. 

Let us also mention here that Lagrangian densities of non-linear extensions of electrodynamics with a logarithmic function of the electromagnetic field strengths are a typical characteristic of QED effective actions. In the classical work by Euler and Heisenberg \cite{Euler}, in which the authors studied electrons in a background set up by a uniform electromagnetic field, a logarithmic term of the field strength came out as an exact $1$-loop correction to the vacuum polarization. Furthermore, some years ago, Volovik \cite{Volovik} has worked out the action for an electromagnetic field that emerges as a collective field in superfluid  $ 
{}^3He - A$; this $4$-dimensional action exhibits a logarithmic factor whose argument is a function of the electromagnetic field strengths \cite{Katsnelson}.

On the other hand, it is worth recalling here that the study of extensions of the Standard Model (SM) such as Lorentz invariance violation and fundamental length, have attracted much attention in the past years \cite{AmelinoCamelia:2002wr,Jacobson:2002hd,Konopka:2002tt,Hossenfelder:2006cw,Nicolini:2008aj}. As is well-known, this is mainly because the SM does not include a quantum theory of gravitation. In fact, the necessity of a new scenario has been suggested to overcome difficulties theoretical in the quantum gravity research. Among these new scenarios, probably the most studied framework are quantum field theories allowing non-commuting position operators 
\cite{Witten:1985cc,Seiberg:1999vs,Douglas:2001ba,Szabo:2001kg,Gomis:2000sp,Bichl:2001nf}, where this non-commutativity is an intrinsic property of space-time. We call attention to the fact that these studies have been achieved by using a star product (Moyal product). More recently, a novel way to formulate noncommutative field theory ( or quantum field theory in the presence of a minimal length) has been proposed in \cite{Euro1, Euro2, Euro3}. Later, it has been shown that this approach can be summarized through the introduction of a new multiplication rule which is known as Voros star-product. Evidently, physics turns out be independent from the choice of the type of product \cite{Jabbari}. With these ideas in mind, in previous studies \cite{Gaete:2011ka,Gaete:2012yu}, we have considered the effect of the spacetime noncommutativity on a physical observable. In fact, we have computed the static potential for axionic electrodynamics both in $(3+1)$ and  $(2+1)$ space-time dimensions, in the presence of a minimal length. The point we wish to emphasize, however, is that our analysis leads to a well-defined noncommutative interaction energy. Indeed, in both cases we have obtained a fully ultraviolet finite static potential. Later, we have extended our analysis  for both Yang-Mills theory and gluodynamics in curved space-time, where we have obtained a string tension finite \cite{Gaete13}. 

Given the outgoing experiments related to photon-photon interaction physics \cite{Davila, Silveira, Kanda}, it is desirable to have some additional understanding of the physical consequences presented by a particular nonlinear electrodynamics, that is, logarithmic electrodynamics. Of special interest will be to study aspects of birefringence  as well as to compute a physical observable. In particular, the static potential between two charges, using the gauge-invariant but path-dependent variables formalism, which is an alternative to the Wilson loop approach. 

Our work is organized according to the following outline: in Section II, we present general aspects of Logarithmic Electrodynamics, show that it yields birefringence and compute the finite electrostatic field-energy of a point-like charge. In Section III, we analyze the interaction energy for a fermion-antifermion pair in usual Logarithmic Electrodynamics and its version in the presence of a minimal length. Finally, in Section IV, we cast our Final Remarks.

\section{The model under consideration}

The model under consideration is described by the Lagrangian density:
\begin{equation}
{\cal L} =  - \beta ^2 \ln \left[ {1 + \frac{{{\cal F}}}{{\beta ^2 }} - \frac{{{\cal G}^2 }}{{2 \beta ^4 }}} \right], \label{LnE05}
\end{equation}
where ${\cal F} = \frac{1}{4}F_{\mu \nu } F^{\mu \nu }$ and $   
{\cal G} = \frac{1}{4}F_{\mu \nu } \tilde F^{\mu \nu }$. As usual, $   
F_{\mu \nu }  = \partial _\mu  A_\nu   - \partial _\nu  A_\mu$ is the electromagnetic field strength tensor and $\tilde F^{\mu \nu }  = \frac{1}{2}\varepsilon ^{\mu \nu \rho \lambda } F_{\rho \lambda }$ is the dual electromagnetic field strength tensor.

The equations of motion follow from Lagrangian density (\ref{LnE05}) read:
\begin{equation}
\partial _\mu  \left[ {\frac{1}{\Gamma }\left( {F^{\mu \nu }  - \frac{1}{{\beta ^2 }}{\cal G}\tilde F^{\mu \nu } } \right)} \right] = 0, \label{LnE10}
\end{equation}
while the Bianchi identities are
\begin{equation}
\partial _\mu  \tilde F^{\mu \nu }  = 0, \label{LnE15}
\end{equation}
where
\begin{equation}
\Gamma  = 1 + \frac{{{\cal F}}}{{\beta ^2 }} - \frac{{{\cal G}^2 }}{{2 \beta ^4 }}. \label{LnE20}
\end{equation}

It follows from the above discussion that Gauss' law takes the form,
\begin{equation}
\nabla  \cdot {\bf D} = 0, \label{LnE20-5}
\end{equation}
where $\bf D$ is given by
\begin{equation}
{\bf D} = \frac{{{\bf E} + \frac{1}{{\beta ^2 }}\left( {{\bf E} \cdot {\bf B}} \right){\bf B}}}{{1 - \frac{{\left( {{\bf E}^2  - {\bf B}^2 } \right)}}{{2\beta ^2 }} - \frac{1}{{2\beta ^4 }}\left( {{\bf E} \cdot {\bf B}} \right)^2 }}. \label{LnE20-10}
\end{equation}
For $J^0 (t,{\bf r}) = e\delta ^{\left( 3 \right)} \left( {\bf r} \right)$, the electric field follows as
\begin{equation}
{\bf E} = \frac{{\beta ^2 }}{Q}\left( {\sqrt {r^4  + \frac{{2Q^2 }}{{\beta ^2 }}}  - r^2 } \right) \hat r,\label{LnE2015}
\end{equation}
or what is the same,
\begin{equation}
{\bf E} = 2Q\frac{1}{{\left( {\sqrt {r^4  + 2\left( {\frac{Q}{{\beta }}} \right)^2 }  + r^2 } \right)}}\hat r. \label{LnE2015b}
\end{equation} $\hat r = \frac{r}{{|r|}}$ and $Q \equiv \frac{e}{{4\pi }}$.  From this expression it should be clear that the electric field of a point like charge is maximum at the origin, ${E_{\max }} = \sqrt 2 \beta$; in the usual Born-Infeld electrodynamics, ${E_{\max }} = \beta$.

In order to write the dynamical equations into a more compact and convenient form,
we shall introduce the vectors ${\bf D} = {{\partial {\cal L}} \mathord{\left/
 {\vphantom {{\partial L} {\partial {\bf E}}}} \right.
 \kern-\nulldelimiterspace} {\partial {\bf E}}}$ and ${\bf H} =  - {{\partial {\cal L}} \mathord{\left/
 {\vphantom {{\partial L} {\partial {\bf B}}}} \right.
 \kern-\nulldelimiterspace} {\partial {\bf B}}}$, in analogy to the electric displacement and magnetic field strength. We then have 
\begin{equation}
{\bf D} = \frac{1}{\Gamma }\left( {{\bf E} + \frac{{{\bf B}\left( {{\bf E} \cdot {\bf B}} \right)}}{{\beta ^2 }}} \right), \label{LnE25}
\end{equation}
and
\begin{equation}
{\bf H} = \frac{1}{\Gamma }\left( {{\bf B} - \frac{{{\bf E}\left( {{\bf E} \cdot {\bf B}} \right)}}{{\beta ^2 }}} \right). \label{LnE30}
\end{equation}
As in the case of usual Born-Infeld electrodynamics, it is worthwhile to invert eq. (\ref{LnE25}), so that we can express ${\bf E}$ in terms of ${\bf D}$ (and ${\bf B}$), in analogy with the Hamiltonian treatment ($\bf E$ could be thought as being the velocity, whereas $\bf D$ plays the role of the momentum). Lengthy algebraic manipulations yield:
\begin{equation}
{\bf E} = \xi {\bf D} + \tilde \xi {\bf B}, \label{ LnE30a}
\end{equation}
where
\begin{equation}
 \xi  = \frac{{ - \beta ^2 \left( {\beta ^2  + {\bf B}^2 } \right)}}{{\left[ {\beta ^2 {\bf D}^2  + \left( {{\bf B} \times {\bf D}} \right)^2 } \right]}}  
+\frac{{\sqrt {\beta ^4 \left( {\beta ^2  + {\bf B}^2 } \right)^2  + \left( {\beta ^2  + {\bf B}^2 } \right)\left( {2\beta ^2  + {\bf D}^2 } \right)\left[ {\beta ^2 {\bf D}^2  + \left( {{\bf B} \times {\bf D}} \right)^2 } \right]} }}{{\left[ {\beta ^2 {\bf D}^2  + \left( {{\bf B} \times {\bf D}} \right)^2 } \right]}} ,   \label{ LnE30b}
 \end{equation}
and
\begin{equation}
\tilde \xi  \equiv \frac{1}{{\sqrt {\beta ^2  + {\bf B}^2 } }}\sqrt { {\bf D}^2 \xi ^2 + 2\beta ^2 \xi  - \left( {2\beta ^2  + {\bf B}^2 } \right)}.  \label{ LnE30c}
\end{equation}
Now, that we have inverted $\bf E$ in terms of $\bf D$, let us also re-express $\bf H$ in terms of $\bf B$ and $\bf D$. We arrive at 
\begin{equation}
{\bf H} = \frac{1}{\xi }\left( {1 + \tilde \xi ^2 } \right){\bf B} + \xi {\bf D}. \label{ LnE30d}
\end{equation}

With this, we can write the corresponding equations of motion as
\begin{equation}
\nabla  \cdot {\bf D} = 0, \  \  \
\frac{{\partial {\bf D}}}{{\partial t}} - \nabla  \times {\bf H} = 0, \label{LnE35a}
\end{equation}
and
\begin{equation}
\nabla  \cdot {\bf B} = 0, \  \  \
\frac{{\partial {\bf B}}}{{\partial t}} + \nabla  \times {\bf E} = 0. \label{LnE35b}\end{equation}
Now, employing (\ref{LnE25}) and (\ref{LnE30}), we then obtain the electric permitivity 
$\varepsilon _{ij}$ and the inverse magnetic permeability 
$\left( {\mu}^{-1}  \right)_{ij}$ tensors of the vacuum, that is,
\begin{equation}
\varepsilon _{ij}  = \frac{1}{\Gamma }\left( {\delta _{ij}  + \frac{1}{{\beta ^2 }}B_i B_j } \right), \ \ \
\left( {\mu}^{-1}  \right)_{ij}  = \frac{1}{\Gamma }\left( {\delta _{ij}  - \frac{1}{{\beta ^2 }}E_i E_j } \right), \label{LnE40}
\end{equation}
with $D_i  = \varepsilon _{ij} E_j$ and $B_i  = \mu _{ij} H_j $.

It is now important to notice that the complicated field problem can be greatly simplified if the equations (\ref{LnE40}) are linearized. As is well-known, this procedure is justified for the description of a weak electromagnetic wave $({\bf E_p}, {\bf B_p})$ propagating in the presence of a strong constant external field $({\bf E_0}, {\bf B_0})$. For computational simplicity our analysis will be developed in the case of a purely magnetic field, that is, ${\bf E_0}=0$. This then implies that  
\begin{equation}
{\bf D} = \frac{1}{{\left( {1 + \frac{{{\bf B}_0^2}}{{{2 \beta ^2}}}} \right)}}\left[ {{{\bf E_p}} + \frac{1}{{{\beta ^2}}}\left( {{{\bf E_p}} \cdot {{\bf B_0}}} \right){{\bf B_0}}} \right], \label{LnE45}
\end{equation}
and
\begin{equation}
{\bf H} = \frac{1}{{\left( {1 + \frac{{{\bf B}_0^2}}{{{2 \beta ^2}}}} \right)}}\left[ {{{\bf B_p}} - \frac{1}{{{\beta ^2}\left( {1 + \frac{{{\bf B}_0^2}}{{{2 \beta ^2}}}} \right)}}\left( {{{\bf B_p}} \cdot {{\bf B}_0}} \right){{\bf B}_0}} \right], \label{LnE50} 
\end{equation}
where we have keep only linear terms in ${\bf E_p}$, ${\bf B_p}$.

Next, without restricting generality we take the $z$ axis as the direction of the magnetic field, ${\bf B_0}  = B_0 {\bf e}_3$, and assuming that the light wave moves along the $x$ axis. We further make a plane wave decomposition for the fields $E_p$ and $B_p$, that is, 
\begin{equation}
{{\bf E_p}}\left( {{\bf x}
,t} \right) = {\bf E}
{e^{ - i\left( {wt - {\bf k} \cdot {\bf x}} \right)}}, \ \ \
{{\bf B_p}}\left( {{\bf x},t} \right) = {\bf B}{e^{ - i\left( {wt - {\bf k} \cdot {\bf x}} \right)}}, \label{LnE55}
\end{equation}
so that the Maxwell equations become
\begin{equation}
\left( {\frac{{{k^2}}}{{{w^2}}} - {\varepsilon _{22}}{\mu _{33}}} \right){E_2} = 0, \label{LnE60a}\end{equation}  
and
\begin{equation}
\left( {\frac{{{k^2}}}{{{w^2}}} - {\varepsilon _{33}}{\mu _{22}}} \right){E_3} = 0. \label{LnE60b}
\end{equation}

As a consequence, we have two different situations: First, if ${\bf E}\ \bot \ {\bf B}_0$ (perpendicular polarization), from (\ref{LnE60b}) $E_3=0$, and from (\ref{LnE60a}) we get $\frac{{{k^2}}}{{{w^2}}} = {\varepsilon _{22}}{\mu _{33}}$. Hence we see that the dispersion relation of the photon takes the form

\begin{equation}
{n_ \bot } = \sqrt {\frac{{1 + {{B_0^2} \mathord{\left/
 {\vphantom {{B_0^2} {{2 \beta ^2}}}} \right.
 \kern-\nulldelimiterspace} {{2 \beta ^2}}}}}{{1 - {{B_0^2} \mathord{\left/
 {\vphantom {{B_0^2} {{2 \beta ^2}}}} \right.
 \kern-\nulldelimiterspace} {{2 \beta ^2}}}}}}. \label{LnE65}
\end{equation}
Second, if ${\bf E}\ || \ {\bf B}_0$ (parallel polarization), from (\ref{LnE60a}) $E_2=0$, and from (\ref{LnE60b}) we get $\frac{{{k^2}}}{{{w^2}}} = {\varepsilon _{33}}{\mu _{22}}$. In this case, the corresponding dispersion relation becomes
\begin{equation}
{n_\parallel } = \sqrt {1 + {{B_0^2} \mathord{\left/
 {\vphantom {{B_0^2} {{\beta ^2}}}} \right.
 \kern-\nulldelimiterspace} {{\beta ^2}}}}.  \label{LnE70}
\end{equation}
This implies that the electromagnetic waves with different polarizations have different velocities or, more precisely, the vacuum birefringence phenomenon is present. Before concluding this section, we should comment on our result. The above result give us an opportunity to compare our result with related nonlinear electrodynamics, that is, Born-Infeld theory. In this case, the theory is written with a square root instead of a logarithm as in (\ref{LnE05}), the phenomenon of birefringence is absent. However,  in the case of a generalized Born-Infeld electrodynamics \cite{Kruglov2}, which contains two different parameters, again the phenomenon of birefringence is present.

Another relevant aspect to compare in both Born-Infeld and logarithmic electrodynamics is the calculation of the finite energy stored in the electrostatic field of a point-like charge; in the case of logarithmic electrodynamics, this field is given by eqs. (\ref{LnE2015}) and (\ref{LnE2015b}). With the general expression for the energy density (the ${\Theta ^{00}}$-component of the energy-momentum tensor, ${\Theta ^{\mu\nu}}$):
\begin{equation}
\Theta _\nu ^\mu   = \frac{{\partial {\cal L}}}{{\partial {\cal F}}}F^{\mu \rho } F_{\nu \rho }  + \frac{{\partial {\cal L}}}{{\partial {\cal G}}}\tilde F^{\mu \rho } F_{\nu \rho }  - \delta _\nu ^\mu  {\cal L}, \label{LnE70b}
\end{equation}
\begin{equation}
\Theta ^{00}  = \frac{1}{\Gamma }{\bf E}^2  + \frac{1}{{\beta ^2 \Gamma }}{\bf E} \cdot {\bf B} + \beta ^2 ln\Gamma,  \label{LnE70c}
\end{equation}
($\Gamma$ is given by eq. (\ref{LnE20})),
in our particular case,
\begin{equation}
\Theta ^{00}  = \frac{{{\bf E}^2 }}{{1 - \frac{{{\bf E}^2 }}{{2\beta ^2 }}}} + \beta ^2 ln\left( {1 - \frac{{{\bf E}^2 }}{{2\beta ^2 }}} \right).  \label{LnE70d}
\end{equation}

From this, we are able to write down the overall (finite) stored electrostatic energy:
\begin{equation}
E_{fin}  = 2\pi Q^{{\raise0.5ex\hbox{$\scriptstyle 3$}
\kern-0.1em/\kern-0.15em
\lower0.25ex\hbox{$\scriptstyle 2$}}} \beta ^{{\raise0.5ex\hbox{$\scriptstyle 1$}
\kern-0.1em/\kern-0.15em
\lower0.25ex\hbox{$\scriptstyle 2$}}} \left( {I_1  + I_2 } \right),   \label{LnE70e}\end{equation}
where
\begin{equation}
I_1  \equiv \int_0^\infty  {d\lambda } \frac{{\left( {\sqrt {2 + \lambda ^2 }  - \lambda } \right)^2 }}{{1 - \frac{1}{2}\left( {\sqrt {2 + \lambda ^2 }  - \lambda } \right)^2 }},  \label{LnE70f}
\end{equation}
and
\begin{equation}
I_2  \equiv \int_0^\infty  {d\lambda } \sqrt \lambda  ln\left[ {1 - \frac{1}{2}\left( {\sqrt {2 + \lambda ^2 }  - \lambda } \right)^2 } \right]. \label{LnE70g}
\end{equation}
Both integrals are finite: $I_1  = 4.157$ and $I_2  =  - 1.385$; from that, we get $E_{fin}$ as given below:
\begin{equation}
E_{fin}  = 0.391\sqrt {e^3 \beta },  \label{LnE70h}
\end{equation}
to be compared with the corresponding value in the usual Born-Infeld case \cite{Birula}:
\begin{equation}
E_{fin}^{BI}  = 1.236\sqrt {e^3 \beta }.  \label{LnE70i}
\end{equation}
By virtue of the logarithmic form of our action (instead of the square root in the Born-Infeld case), it becomes clear why the stored electrostatic energy is smaller, in comparison with the case of Born-Infeld. To get an estimate of the coupling parameter $\beta$, we could identify the maximal electrostatic field,
\begin{equation}
|{E_{\max }}| = \sqrt 2 \beta,  \label{LnE70j}
\end{equation}
with the natural fundamental field that appears in terms of the electron's charge and mass, $m_e$ and the fundamental constants $c$ and $\hbar$:
\begin{equation}
{{\cal E}_{fund}} = \frac{{m_e^2{c^3}}}{{e\hbar }}. \label{LnE70k}
\end{equation}
In natural units ($\hbar$ = $c$ = 1), its value is
\begin{equation}
{{\cal E}_{fund}} = 5.981Me{V^2}, \label{LnE70l}
\end{equation}
which corresponds to $2.590 \times {10^{18}}{\raise0.7ex\hbox{$N$} \!\mathord{\left/
 {\vphantom {N C}}\right.\kern-\nulldelimiterspace}\!\lower0.7ex\hbox{$C$}}$.

If we adopt that $\beta$ is fixed by ${{\cal E}_{fund}}$,
\begin{equation}
\beta  = \frac{{{m^2}}}{{\sqrt 2 e}}, \label{LnE70m}
\end{equation}
then we may compute the total amount of electrostatic energy, $U$, stored in a domain whose radius is the electron's Compton length ($R = \frac{1}{m}$). We get
\begin{equation}
U = 4\pi \int_0^{{\raise0.5ex\hbox{$\scriptstyle 1$}
\kern-0.1em/\kern-0.15em
\lower0.25ex\hbox{$\scriptstyle m$}}} {dr{r^2}} {\Theta ^{00}} = 8.67 \times {10^{ - 4}}{m_e}, \label{LnE70n}
\end{equation}
after we take $\beta$ given by eq. (\ref{LnE70m}) and the integrals $I_1$ and $I_2$ of eqs. (\ref{LnE70f}) and (\ref{LnE70g}) are carried out over the region that corresponds to the electron's Compton length.

At this point, we would like to draw attention the reader's attention to the recent work by Costa et al. \cite{Costa}, where these authors investigate a non-linear gauge-invariant extension of classical electrodynamics, quartic in the field strength (they consider an $F^2$-term) and also attain a finite value for the field energy of a point-like charge.

\section{Interaction energy}

As already stated, our principal purpose is to calculate
explicitly the interaction energy between static point-like sources
for logarithmic electrodynamics. To this end  we will calculate
the expectation value of the energy operator $ H$ in the physical
state $ |\Phi\rangle$, which we will denote by ${ \langle H
\rangle_ \Phi}$. The starting point is the Lagrangian (\ref{LnE05}), that is, 
\begin{equation}
{\cal L} =  - \beta ^2 \ln \left[ {1 + \frac{{{\cal F}}}{{\beta ^2 }} - \frac{{{\cal G}^2 }}{{2\beta ^4 }}} \right]. \label{LnE75}
\end{equation}
Next, we will introduce an auxiliary field $v$ to handle the logarithmic in the Lagrangian density (\ref{LnE75}). Expressed in terms of this field, the corresponding Lagrangian takes the form
\begin{equation}
{\cal L} = \beta ^2  - \beta ^2 \ln \beta ^4  + \beta ^2 \ln v - \frac{v}{{\beta ^2 }}\left[ {1 + \frac{1}{{4\beta ^2 }}F_{\mu \nu }^2  - \frac{1}{{32\beta ^4 }}\left( {F_{\mu \nu } \tilde F^{\mu \nu } } \right)^2 } \right]. \label{LnE80}
\end{equation}

With this in hand, the canonical momenta are $   
\Pi ^\mu   =  - \frac{v}{{\beta ^4 }}\left( {F^{0\mu }  - \frac{v}{{4\beta ^2 }}F_{\alpha \beta } \tilde F^{\alpha \beta } \tilde F^{0\mu } } \right)$, and one
immediately identifies the two primary constraints $ \Pi ^0  = 0 $
and $ p \equiv \frac{{\partial {\cal L}}}{{\partial v}} = 0$. The
canonical Hamiltonian of the model can be worked out as usual and
is given by the expression
\begin{eqnarray}
H_C  &=& 
\int {d^3 x} \left\{ {\Pi _i \partial ^i A_0  + \frac{{\beta ^4 }}{{2v}}{\bf \Pi} ^2  + \frac{v}{{\beta ^2 }}\left( {1 + \frac{{{\bf B}^2 }}{{2\beta ^2 }}} \right) - \frac{{\beta ^2 }}{{2v}}\frac{{\left( {{\bf \Pi}  \cdot {\bf B}^2 } \right)}}{{\left( {1 + \frac{{{\bf B}^2 }}{{\beta ^2 }}} \right)}}}  + \beta ^2  - \beta ^2 \ln \beta ^4  + \beta ^2 \ln v \right\}. \label{LnE85}
\end{eqnarray}
Now, requiring the primary constraint $\Pi _0 $ to be preserved in time
yields the secondary constraint (Gauss' law) $\Gamma _1 \left( x
\right) \equiv\partial _i \Pi ^i =0$. Similarly for the
constraint $p$, we get the auxiliary field $v$ as
\begin{equation}
v = \frac{{\beta ^4 }}2{{\left( {1 + \frac{{{\bf B}^2 }}{{2\beta ^2 }}} \right)}}\left\{ {1 + \sqrt {1 + \frac{2}{{\beta ^2 }}\left( {1 + \frac{{{\bf B}^2 }}{{2\beta ^2 }}} \right)\left[ {{\bf \Pi} ^2  - \frac{{\left( {{\bf B} \cdot {\bf \Pi} } \right)^2 }}{{\beta ^2 \left( {1 + \frac{{{\bf B}^2 }}{{\beta ^2 }}} \right)}}} \right]} } \right\}. \label{LnE90}
\end{equation}

The extended Hamiltonian that generates translations in time then
reads $H = H_C  + \int {d^3 } x\left( {c_0 (x)\Pi_0 (x) + c_1
(x)\Gamma _1 (x)} \right)$, where $c_0(x)$ and $c_1(x)$ are
Lagrange multipliers. In addition, neither $ A_0 \left( x \right)$
nor $ \Pi _0 \left( x \right)$ are of interest in describing the
system and may be discarded from the theory. Thus we are left with
the following expression for the Hamiltonian
\begin{eqnarray}
H &=& \int {d^3 x} \Biggl[c^{\prime}\left( x \right)\partial _i \Pi ^i  + {{\bf \Pi} ^2 }\frac{{\left( {1 + \frac{{{\bf B}^2 }}{{2\beta ^2 }}} \right)}}{{\left\{ {1 + \sqrt {1 + \frac{2}{{\beta ^2 }}\left( {1 + \frac{{{\bf B}^2 }}{{2\beta ^2 }}} \right)\left[ {{\bf \Pi} ^2  - \frac{{\left( {{\bf B} \cdot {\bf \Pi} } \right)^2 }}{{\beta ^2 \left( {1 + \frac{{{\bf B}^2 }}{{\beta ^2 }}} \right)}}} \right]} } \right\}}}  \nonumber\\ 
&+& \frac{{{\bf B}^2 }}{4}\frac{1}{{\left( {1 + \frac{{{\bf B}^2 }}{{2\beta ^2 }}} \right)}}\left\{ {1 + \sqrt {1 + \frac{2}{{\beta ^2 }}\left( {1 + \frac{{{\bf B}^2 }}{{2\beta ^2 }}} \right)\left[ {{\bf \Pi} ^2  - \frac{{\left( {{\bf B} \cdot {\bf \Pi} } \right)^2 }}{{\beta ^2 \left( {1 + \frac{{{\bf B}^2 }}{{\beta ^2 }}} \right)}}} \right]} } \right\} \nonumber\\ 
&-& \frac{{\left( {{\bf B} \cdot {\bf \Pi} } \right)^2 }}{{4\beta ^2 \left( {1 + \frac{{{\bf B}^2 }}{{\beta ^2 }}} \right)}}\frac{{\left( {1 + \frac{{{\bf B}^2 }}{{2\beta ^2 }}} \right)}}{{\left\{ {1 + \sqrt {1 + \frac{2}{{\beta ^2 }}\left( {1 + \frac{{{\bf B}^2 }}{{2\beta ^2 }}} \right)\left[ {{\bf \Pi} ^2  - \frac{{\left( {{\bf B} \cdot {\bf \Pi} } \right)^2 }}{{\beta ^2 \left( {1 + \frac{{{\bf B}^2 }}{{\beta ^2 }}} \right)}}} \right]} } \right\}}} \nonumber\\ 
&+& \frac{{\beta ^2 }}2{{\left( {1 + \frac{{{\bf B}^2 }}{{2\beta ^2 }}} \right)}}\left\{ {1 + \sqrt {1 + \frac{2}{{\beta ^2 }}\left( {1 + \frac{{{\bf B}^2 }}{{2\beta ^2 }}} \right)\left[ {{\bf \Pi} ^2  - \frac{{\left( {{\bf B} \cdot {\bf \Pi} } \right)^2 }}{{\beta ^2 \left( {1 + \frac{{{\bf B}^2 }}{{\beta ^2 }}} \right)}}} \right]} } \right\} \nonumber\\ 
&-& \beta ^2  + \beta ^2 \ln \beta ^4 - \beta ^2 \ln \left[ {\frac{{\beta ^4 }}2{{\left( {1 + \frac{{{\bf B}^2 }}{{2\beta ^2 }}} \right)}}} \right] \nonumber\\
&-& \beta ^2 \ln \left\{ {1 + \sqrt {1 + \frac{2}{{\beta ^2 }}\left( {1 + \frac{{{\bf B}^2 }}{{2\beta ^2 }}} \right)\left[ {{\bf \Pi} ^2  - \frac{{\left( {{\bf B} \cdot {\bf \Pi} } \right)^2 }}{{\beta ^2 \left( {1 + \frac{{{\bf B}^2 }}{{\beta ^2 }}} \right)}}} \right]} } \right\}\Biggr]\ , \label{LnE95}
\end{eqnarray}
where $ c^{\prime }(x)=c_{1}(x)-A_{0}(x)$.

Next, since there is one first class constraint $\Gamma_{1}(x)$ (Gauss' law),
we choose one gauge fixing condition that will make the full set of
constraints becomes second class. We choose the gauge fixing condition
to correspond to
\begin{equation}
\Gamma _2 \left( x \right) \equiv \int\limits_{C_{\xi x} } {dz^\nu
} A_\nu \left( z \right) \equiv \int\limits_0^1 {d\lambda x^i }
A_i \left( {\lambda x} \right) = 0. \label{LnE100}
\end{equation}
where  $\lambda$ $(0\leq \lambda\leq1)$ is the parameter describing
the spacelike straight path $ x^i = \xi ^i  + \lambda \left( {x -
\xi } \right)^i $, and $ \xi $ is a fixed point (reference point).
There is no essential loss of generality if we restrict our
considerations to $ \xi ^i=0 $. The choice (\ref{LnE100}) leads to
the Poincar\'e gauge \cite{PG1,PG2}. As a consequence, we can now
write down the only nonvanishing Dirac bracket for the canonical
variables
\begin{equation}
\left\{ {A_i \left( x \right),\Pi ^j \left( y \right)} \right\}^ *
= \delta _i^j \delta ^{\left( 3 \right)} \left( {x - y} \right) -
\partial _i^x \int\limits_0^1 {d\lambda x^i } \delta ^{\left( 3
\right)} \left( {\lambda x - y} \right). \label{LnE105}
\end{equation}

We are now in a position to compute the potential energy for
static charges in this theory. To do this, we consider will use the
gauge-invariant scalar potential which is given by
\begin{equation}
V \equiv e\left( {{\cal A}_0 \left( {\bf 0} \right) - {\cal A}_0 \left( {\bf L} \right)} \right), \label{LnE110}
\end{equation}
where the physical scalar potential is given by
\begin{equation}
{\cal A}_0 (t,{\bf r}) = \int_0^1 {d\lambda } r^i E_i (t,\lambda
{\bf r}). \label{LnE115}
\end{equation}
This equation follows from the vector gauge-invariant field expression
\begin{equation}
{\cal A}_\mu  (x) \equiv A_\mu  \left( x \right) + \partial _\mu  \left( { - \int_\xi ^x {dz^\mu  A_\mu  \left( z \right)} } \right), \label{LnE120}
\end{equation}
where the line integral is along a spacelike path from the point $\xi$ to $x$, on a fixed slice time. It should be noted that the gauge-invariant variables (\ref{LnE120}) commute with the sole first constraint (GaussÕ law), showing in this way that these fields are physical variables.

Having made these observations, we see that GaussÕs law for the present theory (obtained from the Hamiltonian formulation above) leads to $\partial _i \Pi ^i  = J^0$, where we have included the external current $J^0$ to represent the presence of two opposite charges. For $J^0 \left( {t,{\bf x}} \right) = Q\delta ^{\left( 3 \right)} \left( {\bf x} \right)$, the electric field then becomes
\begin{equation}
{\bf E} = \frac{Q}{{2\pi }}\frac{1}{{\left( {\sqrt {r^4  + 2\left( {\frac{Q}{{\beta 4\pi }}} \right)^2 }  + r^2 } \right)}}\hat r. \label{LnE125}
\end{equation}

As a consequence, equation (\ref{LnE115}) becomes
 \begin{equation}
{{\cal A}_0} =  - \frac{Q}{{2\pi }}\left\{ {\frac{{2\sqrt 2 \beta }}{Q} \hspace{1mm} _2{F_1}\left( { - {\textstyle{1 \over 2}},{\textstyle{1 \over 4}},{\textstyle{5 \over 4}}, - {\textstyle{{8{\pi ^2}{\beta ^2}} \over {{Q^2}}}}{r^4}} \right) r - \frac{{8{\pi ^2}{\beta ^2}}}{{3{Q^2}}}{r^3}} \right\}, \label{LnE125c}
\end{equation}
where $_2{F_1}$ is the hypergeometric function.
 In terms of $ {\cal A}_0 \left( {t,{\bf r}} \right) $, the
potential for a pair of static point-like opposite charges located
at $\bf 0$ and $\bf L$, is given by 
\begin{eqnarray}
V &\equiv& Q \left( {{\cal A}_0 \left( {\bf 0} \right) - {\cal A}_0
\left( {\bf L} \right)} \right) \nonumber\\
&= & \frac{Q^2}{{2\pi }}\left\{ {\frac{{2\sqrt 2 \beta }}{Q} \hspace{1mm} _2{F_1}\left( { - {\textstyle{1 \over 2}},{\textstyle{1 \over 4}},{\textstyle{5 \over 4}}, - {\textstyle{{8{\pi ^2}{\beta ^2}} \over {{Q^2}}}}{L^4}} \right) L - \frac{{8{\pi ^2}{\beta ^2}}}{{3{Q^2}}}{L^3}} \right\},
\label{LnE125d}
\end{eqnarray} 
with $ L = |{\bf L}|$.

The above analysis give us an opportunity to compare logarithmic electrodynamics with related Born-Infeld electrodynamics. In this case, the electric field is given by
 \begin{equation}
{\bf E} = \frac{Q}{{4\pi }}\frac{1}{{\sqrt {{r^4} + {\textstyle{{{Q^2}} \over {{{\left( {4\pi \beta } \right)}^2}}}}} }}\hat r,\label{LnE125e}
 \end{equation}
from which follows that 
\begin{equation}
V = Q\beta \hspace{1mm}_2{F_1}\left( {{\textstyle{1 \over 4}},{\textstyle{1 \over 2}},{\textstyle{5 \over 4}}, - {\textstyle{{4\pi \beta } \over Q}}{L^2}} \right)L. \label{LnE125f}
\end{equation} 
The plot of eqs. (\ref{LnE125c}) and (\ref{LnE125f}) is showed in Fig. $1$.

We further note that the scalar potential for logarithmic electrodynamics, at leading order in $\beta$, takes the form
\begin{equation}
{\cal A}_0 \left( {t,{\bf r}} \right) =  - \frac{Q}{{8\pi r}}\int_0^1 {d\lambda } \left\{ {\frac{1}{{\lambda ^2 }} - \frac{1}{4}\frac{{a^4 }}{{\lambda ^6 }}} \right\}, \label{LnE130}
\end{equation}
where $ a^4  \equiv \frac{{\rho _0^2 }}{{r^4 }} = \frac{{Q^2
}}{{8\beta ^2 \pi ^2 r^4 }} $. In this way, by employing Eq. (\ref{LnE130}), the
potential for a pair of static point-like opposite charges located
at $\bf 0$ and $\bf L$, is given by
\begin{equation}
V \equiv Q \left( {{\cal A}_0 \left( {\bf 0} \right) - {\cal A}_0
\left( {\bf L} \right)} \right) =  - \frac{{Q^2 }}{{8\pi
}}\frac{1}{L}\left( {1 - \frac{{e^2 }}{{160\pi ^2 \beta ^2
}}\frac{1}{{L^4 }}} \right). \label{LnE130b}
\end{equation}
Thus, to ${\cal O}\left( {\frac{1}{{\beta ^2 }}} \right)$, logarithmic
electrodynamics displays a marked qualitative departure from the
usual Maxwell theory. More importantly, this is exactly the profile obtained
for Born-Infeld electrodynamics. Accordingly, logarithmic electrodynamics 
also has a rich structure reflected by its long-range correction to the 
Coulomb potential. 
\begin{figure}[h]
\begin{center}
\includegraphics[scale=1.4]{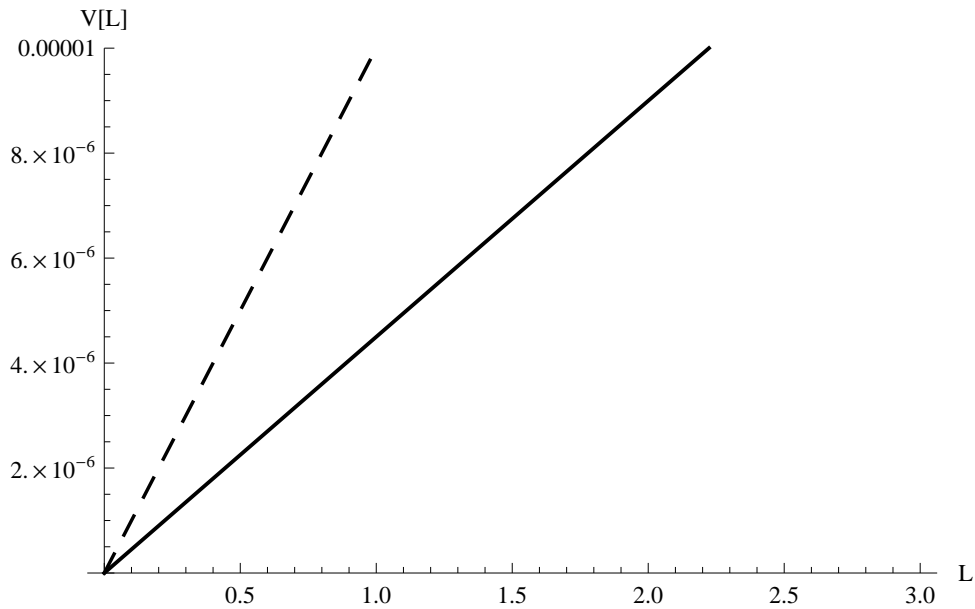}
\end{center}
\caption{\small Shape of the potential, Eqs.  (\ref{LnE125c})(Solid line) and (\ref{LnE125f})(Dashed line)
\label{fig 1}}
\end{figure}

At this point an interesting matter comes out. Although logarithmic electrodynamics has a finite electric field at the origin, the interaction energy between two test charges at leading order in $\beta$ is not finite at the origin. In view of this situation, we now proceed to examine the behavior of logarithmic electrodynamics defined in a non-commutative geometry, along the lines of references \cite{Gaete:2011ka,Gaete:2012yu}. Basically, our goal is to explore the behavior of the interaction energy at short distances. In this case, Gauss' law reads
\begin{equation}
\partial _i \Pi ^i  = e \ e^{ - \theta \nabla ^2 } \delta ^{\left( 3 \right)} \left( {\bf x} \right). \label{LnE135}
\end{equation}
This then implies that
\begin{equation}
\Pi ^i  = -\frac{{2e}}{{\sqrt \pi  }}\frac{{\hat r^i }}{{r^2 }}\gamma \left( {{\raise0.5ex\hbox{$\scriptstyle 3$}
\kern-0.1em/\kern-0.15em
\lower0.25ex\hbox{$\scriptstyle 2$}},{\raise0.5ex\hbox{$\scriptstyle {r^2 }$}
\kern-0.1em/\kern-0.15em
\lower0.25ex\hbox{$\scriptstyle {4\theta }$}}} \right), \label{LnE140}
\end{equation}
with $r = |{\bf r}|$. Here $\gamma \left( {{\raise0.5ex\hbox{$\scriptstyle 3$}
\kern-0.1em/\kern-0.15em
\lower0.25ex\hbox{$\scriptstyle 2$}},{\raise0.5ex\hbox{$\scriptstyle {r^2 }$}
\kern-0.1em/\kern-0.15em
\lower0.25ex\hbox{$\scriptstyle {4\theta }$}}} \right)$ is the lower incomplete Gamma function defined by the following integral representation
\begin{equation}
\gamma \left( {{\raise0.5ex\hbox{$\scriptstyle a$}
\kern-0.1em/\kern-0.15em
\lower0.25ex\hbox{$\scriptstyle b$}},x} \right) \equiv \int_0^x {\frac{{du}}{u}} u^{{\raise0.5ex\hbox{$\scriptstyle a$}
\kern-0.1em/\kern-0.15em
\lower0.25ex\hbox{$\scriptstyle b$}}} e^{ - u}. \label{LnE145}
\end{equation}
 Next, from expression for the electric field, we have
\begin{equation}
E_i  = e\frac{1}{{\left[ {1 + \sqrt {1 + {\textstyle{2 \over {\beta ^2 }}}{\bf \Pi} ^2 } } \right]}}\partial _i \left( { - \frac{{e^{ - \theta \nabla ^2 } \delta ^{\left( 3 \right)} \left( {\bf x} \right)}}{{\nabla ^2 }}} \right), \label{LnE150}
\end{equation}
in this last line we have considered the static case (${\bf B}=0$). At leading order in $\beta$, the electric field follows as 
\begin{equation}
E_i  = \frac{e}{2}\left( {1 - \frac{{{\bf \Pi} ^2 }}{{2\beta ^2 }}} \right)\partial _i \left( { - \frac{{e^{ - \theta \nabla ^2 } \delta ^{\left( 3 \right)} \left( {\bf x} \right)}}{{\nabla ^2 }}} \right), \label{LnE155}
\end{equation}
where $\bf \Pi$ is given by expression (\ref{LnE140}).

Using this result, the physical scalar potential, Eq. (\ref{LnE115}), takes the form
\begin{equation}
{\cal A}_0 \left( {t,{\bf r}} \right) = \frac{e}{2}\int_0^1 {d\lambda \ r^i \partial _i^{\lambda r} } \tilde G\left( {\lambda {\bf r}} \right) - \frac{e}{{4\beta ^2 }}\int_0^1 {d\lambda } \ {\bf \Pi} ^2 \left( {\lambda {\bf r}} \right)r^i\partial _i^{\lambda {\bf r}} \tilde G\left( {\lambda {\bf r}} \right), \label{LnE160}
\end{equation}
where $   
\tilde G({\bf r}) = \frac{1}{{4\pi ^{{\raise0.5ex\hbox{$\scriptstyle 3$}
\kern-0.1em/\kern-0.15em
\lower0.25ex\hbox{$\scriptstyle 2$}}} }}\frac{1}{r}\gamma \left( {{\raise0.5ex\hbox{$\scriptstyle 1$}
\kern-0.1em/\kern-0.15em
\lower0.25ex\hbox{$\scriptstyle 2$}},{\raise0.5ex\hbox{$\scriptstyle {r^2 }$}
\kern-0.1em/\kern-0.15em
\lower0.25ex\hbox{$\scriptstyle {4\theta }$}}} \right)$.
By employing Eq. (\ref{LnE140}) we can reduce Eq. (\ref{LnE160}) to
\begin{equation}
{\cal A}_0 ({\bf r}) = \frac{e}{{8\pi ^{{\raise0.5ex\hbox{$\scriptstyle 3$}
\kern-0.1em/\kern-0.15em
\lower0.25ex\hbox{$\scriptstyle 2$}}} }}\frac{1}{r}\gamma \left( {{\raise0.5ex\hbox{$\scriptstyle 1$}
\kern-0.1em/\kern-0.15em
\lower0.25ex\hbox{$\scriptstyle 2$}},{\raise0.5ex\hbox{$\scriptstyle {r^2 }$}
\kern-0.1em/\kern-0.15em
\lower0.25ex\hbox{$\scriptstyle {4\theta }$}}} \right) + \frac{{2e^3 }}{{\left( \pi  \right)^{{\raise0.5ex\hbox{$\scriptstyle 3$}
\kern-0.1em/\kern-0.15em
\lower0.25ex\hbox{$\scriptstyle 2$}}} \beta ^2 }}\hat r^i \int_{\bf 0}^{\bf r} {dy^i } \frac{1}{{y^6 }}\gamma ^3 \left( {{\raise0.5ex\hbox{$\scriptstyle 3$}
\kern-0.1em/\kern-0.15em
\lower0.25ex\hbox{$\scriptstyle 2$}},{\raise0.5ex\hbox{$\scriptstyle {y^2 }$}
\kern-0.1em/\kern-0.15em
\lower0.25ex\hbox{$\scriptstyle {4\theta }$}}} \right). \label{LnE160}
\end{equation}

Finally, replacing this result in (\ref{LnE110}), the potential for a pair of point-like opposite charges $e$, located at $\bf 0$ and $\bf L$, takes the form
\begin{equation}
V =  - \frac{{e^2 }}{{8\pi ^{{\raise0.5ex\hbox{$\scriptstyle 3$}
\kern-0.1em/\kern-0.15em
\lower0.25ex\hbox{$\scriptstyle 2$}}} }}\frac{1}{L}\left[ {\gamma \left( {{\raise0.5ex\hbox{$\scriptstyle 1$}
\kern-0.1em/\kern-0.15em
\lower0.25ex\hbox{$\scriptstyle 2$}},{\raise0.5ex\hbox{$\scriptstyle {L^2 }$}
\kern-0.1em/\kern-0.15em
\lower0.25ex\hbox{$\scriptstyle {4\theta }$}}} \right) + \frac{{16e^2 }}{{\beta ^2 }}L\ \hat r^i \int_{\bf 0}^{\bf L} {dy^i } \frac{1}{{y^6 }}\gamma ^3 \left( {{\raise0.5ex\hbox{$\scriptstyle 3$}
\kern-0.1em/\kern-0.15em
\lower0.25ex\hbox{$\scriptstyle 2$}},{\raise0.5ex\hbox{$\scriptstyle {y^2 }$}
\kern-0.1em/\kern-0.15em
\lower0.25ex\hbox{$\scriptstyle {4\theta }$}}} \right)} \right]. \label{LnE165}
\end{equation}
One immediately observes that the introduction of the non-commutative space induces a finite static potential for $L \to 0$ (See Fig. 2). This then implies that the self-energy and the electromagnetic mass of a point-like particle are finite in this version non-commutative of logarithmic electrodynamics. It is also important to note that in the limit  $\theta \to 0$, we recover our previous result (\ref{LnE130}).
\begin{figure}[h]
\begin{center}
\includegraphics[scale=1.4]{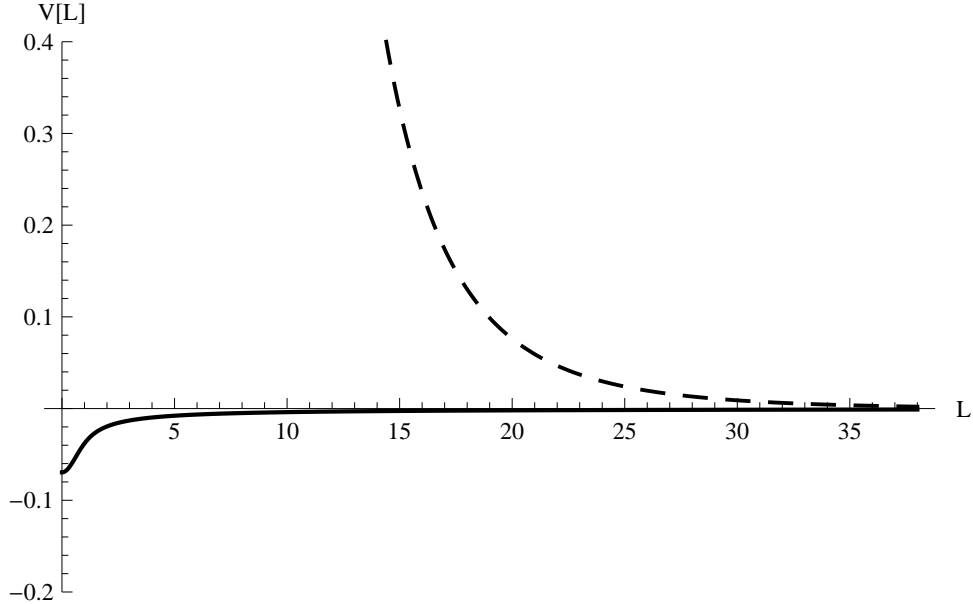}
\end{center}
\caption{\small Shape of the potential, Eqs.  (\ref{LnE165})(Solid line) and (\ref{LnE130b})(Dashed line)
\label{fig 2}} 
\end{figure}

\section{Final Remarks}

In summary, within the gauge-invariant but path-dependent variables formalism, we have considered the confinement versus screening issue for logarithmic electrodynamics. Once again, a correct identification of physical degrees of freedom has been fundamental for understanding the physics hidden in gauge theories. We should highlight the different behaviors of the potentials associated to each of the models. In the logarithmic electrodynamics case, the static potential profile is similar to that encountered in Born-Infeld electrodynamics. Interestingly enough, its non-commutative version displays an ultraviolet finite static potential. The above analysis reveals the key role played by the new quantum of length in our analysis. In a general perspective, the benefit of considering the present approach is to provide unifications among different models, as well as exploiting the equivalence in explicit calculations, as we have illustrated in the
course of this work.

Finally, we should not conceive the electron simply as an electric monopole. The electron's electric dipole moment has recently been re-measured and its upper bound has been improved by a factor around $12$ \cite {Baron}:
\begin{equation}
d_e  \le 10^{ - 29} e. cm  \label{LnE166}
\end{equation}
This means that, at distances of the order of $10^{ - 29}$ cm, one can think of the electron's charge being non-symmetrically distributed around the electron's spin. Moreover, the electron is also a magnetic dipole. So, a very natural path to go deeper into the study of logarithmic electrodynamics would be the investigation of the electron's magnetic dipole moment in terms of the magnetic field induced, through the non-linearity, by the electrostatic field of eqs. (\ref{LnE2015}) and (\ref{LnE2015b}). A step towards this investigation was given in the paper of Ref. \cite{Vellozo}, where the authors attempt at an understanding of the electron's magnetic moment as a non-linear effect induced by its own electrostatic field in the usual Born-Infeld scenario. We shall now focus on the electron's electric and magnetic dipoles in the framework of logarithmic electrodynamics. The results of our pursuit shall be reported elsewhere.

\begin{acknowledgments}
This work was partially supported by Fondecyt (Chile) Grant 1130426.
\end{acknowledgments}

\end{document}